\documentclass{llncs}
\usepackage{epsfig}
\usepackage{amsmath}
\usepackage{amssymb}

\begin{document}
\title{Comparison of Two Numerical Methods for Computation of American Type of the
Floating Strike Asian Option}
\author{J. D. Kandilarov$^1$, D. \v{S}ev\v{c}ovi\v{c}$^2$ }
\institute{ $^1$
Department of Mathematics, University of Rousse,
\\
 e-mail: ukandilarov@uni-ruse.bg , \\
$^2$Department of Applied Mathematics and Statistics, Comenius University\\
e-mail: sevcovic@fmph.uniba.sk }
\date{ }
\maketitle

\begin{abstract}

We present a numerical approach for solving the free boundary problem for the
Black-Scholes equation for pricing American style of floating strike
Asian options. A fixed domain transformation of the free boundary
problem into a parabolic equation defined on a fixed spatial domain
is performed. As a result a nonlinear time-dependent term is
involved in the resulting equation. Two new numerical algorithms are
proposed.  In the first algorithm a predictor-corrector scheme is used.
The second one is based on the Newton method. Computational experiments,
confirming the accuracy of the algorithms are presented and discussed.

\end{abstract}

\section{Introduction}



In this paper we consider the problem of pricing American style
Asian options, analyzed by Bokes and the second author in \cite{BokasSev} (see also \cite{SevcovEhrh}). Asian
options belong to the group of the so-called path-dependent options.
Their pay-off diagrams depend on the spot value of the underlying asset
during the whole or some part(s) of the life span of the option.
Among path-dependent options, Asian option depend is on the arithmetic or geometric average of
spot prices of the underlying asset.
During the last decade, the problem of solving the American option problem numerically has been subject for intensive research
\cite{BokasSev,Kwok,NielSkav,Sev,StaSev} (see also \cite{SevcovEhrh} for overview). A comprehensive
introduction to this topic can be found in \cite{Kwok}. Comparison of various analytical and numerical approximation methods of calculation of  the early
exercise boundary a position of the American put option paying zero dividends
is given in \cite{LauSev}. An improvement of Han and Wu's algorithm
\cite{HanWu} is described in \cite{Tangman}. Our goal is to propose and investigate two front-fixing numerical algorithms for solving free boundary value problems. The front-fixing method has been successfully applied to a wide range of applied problems arising from
physics and engineering, cf. \cite{Gupta,MoyScar} and references
therein. The basic idea is to remove the moving boundary by a
transformation of the involved variables. Transformation techniques were used in the analysis and numerical
computation of the early exercise boundary in the context of American style of
vanilla options \cite{Sev} as well as Asian floating strike options  \cite{BokasSev,SevcovEhrh,SevTak}. In comparison to the existing computational method \cite{BokasSev} we do not replace the algebraic constraint  by its equivalent integral form (see \cite{BokasSev,SevTak} for details) which is computationally more involved. In this paper we solve the corresponding parabolic equation with an algebraic constraint directly as it was proposed in \cite{SevcovEhrh}. The approach presented in \cite{SevcovEhrh} however suffered from the necessity of taking very small time discretization steps. Here we overcome this difficulty by proposing two new numerical approximation algorithms (see Section 4). They are based on the novel technique proposed by the first author and Valkov in \cite{KanVal}. We extend this approach for  American style of Asian options. In Section 5, a numerical example illustrating the capability of our algorithms are discussed.

\section{The Free Boundary Problem}
Following the classical Black-Scholes theory, the second author and Bokes
\cite{BokasSev} analyzed the problem of pricing Asian
options with arithmetically averaged strike price by means of a solution to a parabolic PDE with a free boundary $S_f(t,A)$:
\begin{eqnarray}\label{eq.1}
  &&\frac{\partial V}{\partial t}+
  \frac{\sigma^2}{2}S^2\frac{\partial^2 V}{\partial S^2}
  +(r-q)S\frac{\partial V}{\partial S}
  +\frac{S-A}{t}\frac{\partial V}{\partial A} - rV =0,
\end{eqnarray}
$0<t<T,\;\; 0<S<S_f(t,A),$ satisfying the boundary conditions
\begin{eqnarray}
 V(t,0,A)=0,&&  \;\; \hbox{for any $A>0$ and $0<t<T$,} \label{eq.2} \\
  \frac{\partial V}{\partial S}(t,S_f(t,A),A)=1,&& V(t,S_f(t,A),A)=S_f(t,A)-A,\label{eq.3}
\end{eqnarray}
and the terminal condition (terminal pay-off condition) at the maturity
time $T$:
\begin{eqnarray}\label{eq.4}
V(T,S,A)=\max(S-A,0), \;\;\;\; S, A >0\;.
\end{eqnarray}
Here $S>0$ is the stock price, $A>0$ is the averaged strike price,
$r>0$ is the risk-free interest rate, $q>0$ is a continuous dividend
rate and $\sigma > 0$ is the volatility of the underlying asset
returns. The arithmetically averaged price $A=A_t$ calculated from
the price path $\{S_u, u\in [0,T]\}$ at the time $t$ is defined as
$A_t=\frac1t\int_0^t{S_u}\,du$. For floating strike Asian options,
it is well known (see e.g. \cite{Kwok,DaiKwok,BokasSev}) that one
can perform a dimension reduction by introducing a new time variable
$\tau=T-t$ and a similarity variable $x$ defined as: $x=A/S, \qquad
W(x,\tau)=V(t,S,A)/A$. The spatial domain for the reduced equation
is given by $1/\rho(\tau)<x<\infty$, $\tau \in (0,T)$,
$\rho(\tau)=S_f(T-\tau, A)/A$. Following
(\cite{Sev,StaSev,BokasSev}), we can apply the Landau fixed domain
transformation for the free boundary problem by introducing a new
state variable $\xi$ and an auxiliary function
$\Pi(\xi,\tau)=W(x,\tau)+x\frac{\partial W}{\partial x}(x,\tau)$,
representing a synthetic portfolio. Here $
\xi=\ln\left({\rho(\tau)x}\right)$. In  \cite{BokasSev,Sev,StaSev}
it is shown that under suitable regularity assumptions on the input
data the free boundary problem (\ref{eq.1})--(\ref{eq.4}) can be
transformed into the initial boundary value problem for parabolic
PDE:
\begin{eqnarray}
&& \frac{\partial \Pi}{\partial \tau}+\alpha(\xi,\tau)\frac{\partial
\Pi}{\partial \xi}- \frac{\sigma^2}{2}\frac{\partial^2 \Pi}
{\partial \xi^2}+\beta(\xi,\tau)\Pi= 0,\;\;\;\;\xi>0,\;\;\tau \in (0,T),\label{eq.7} \\
&& \Pi(0,\tau)=-1, \Pi(\infty, \tau)=0, \ \
  \Pi(\xi,0)=\left\{
           \begin{array}{cc}
             -1, &
 \mbox{for} \; \xi< \ln{\rho(0)},
\\
             0, & \mbox{otherwise}.
           \end{array}
         \right.
\label{eq.9}
\end{eqnarray}
The coefficients $\alpha$ and $\beta$ are defined as follows:
\begin{equation}
\alpha(\xi,\tau)=\frac{\dot{\rho}(\tau)}{\rho(\tau)}+r-q- \frac{\sigma^2}{2}-
\frac{\rho(\tau) e^{-\xi} -1 }{T-\tau}, \quad
\beta(\xi,\tau)=r + \frac{1}{T-\tau}.
\label{eq.10}
\end{equation}
According to  \cite{BokasSev} the free boundary function $\rho(\tau)$ and the solution $\Pi$ should fulfill the constraint:
\begin{equation}\label{eq.13}
\rho(\tau)=\frac{1+r(T-\tau)+\frac{\sigma^2}{2}(T-\tau)\frac{\partial
\Pi}{\partial \xi}(0,\tau)}{1+q(T-\tau)}, \quad
\rho(0)=\max\left(\frac{1+rT}{1+qT},1\right).
\end{equation}
As for derivation of the initial free boundary position $\rho(0)$ in  (\ref{eq.13}) we refer to \cite{BokasSev} or \cite{Kwok,DaiKwok}.
A solution $\Pi$ to the problem (\ref{eq.7})-(\ref{eq.13}) is
continuous for $t>0$. The discontinuity appears only at the point
$P^{\star}=(\ln(\rho(0)),0)$. The derivatives of the solution exist
and are sufficiently smooth in $[0,L]\times [0,T)$, outside of a
small neighbourhood of $P^{\star}$. Another important fact to emphasize is that
for times $t\to 0^+$ (i.e. when $\tau \rightarrow T$) the coefficients $\alpha,\beta$ become unbounded.

\section{Finite Difference Schemes}
In order to solve the problem (\ref{eq.7})-(\ref{eq.13})
numerically, we introduce $L$ which is sufficiently large upper limit of values of the $\xi$ variable
(a safe choice is to take $L$ is equal to five times $\ln(\rho(0))$), where
we prescribe $ \Pi(L, \tau ) = 0$. Next, for given positive integers
$N$ and $M$ we define the uniform meshes: $ {\overline {\omega}}_{h}
= \{ 0 \} \cup \{ L \} \cup \omega_{h}, \; \;
  \omega_{h} = \{ \xi_{i} = i h , \; i=1 , \dots , (N-1), \; h = L/N \}$ and
$ {\overline {\omega}}_{k}  = \{ 0 \} \cup  \{ T \} \cup \omega_{k}
    , \; \;
  \omega_{k} = \{ \tau_{j} = jk, \; j=1 , \dots , (M-1) , \; k = T/M\} $.
Our goal is to define a finite difference method which is suitable for
computing $y_{i}^{j} \approx \Pi(\xi_{i},\tau_{j})$ for
$(\xi_{i},\tau_{j}) \in \omega_{h} \times \omega_{k}$ and
associated front position $ z^{j} \approx \rho (\tau_{j})$ for
$\tau_{j}\in \omega_{k}$. The implicit difference scheme has the
following form:
\begin{eqnarray}\label{eq.14}
  \frac{ y_{ i }^{ j+1 } - y_{ i }^{ j }  }
{ k } &+& \alpha_{i}^{j+1} \frac{ y_{ i+1 }^{ j+1 } - y_{ i-1 }^{
j+1 } } { 2h} - \frac{ \sigma^{2} }{2} \frac{ y_{ i+1 }^{ j+1 } - 2
y_{ i }^{j+1 } + y_{ i-1 }^{ j+1 }   } { h^{2} }+\beta^{j+1} y_{ i }^{ j+1 }= 0,\\
&&  y_{ 0}^{ j+1 } = - 1 , \; \; y_{ N }^{ j+1 } = 0 ; \; \; y_{ i
}^{ 0 } =
   \left\{
     \begin{array}{rl}
      -1, &  \mbox{for}\;\; \xi_i \leq \ln(\rho(0)),\\
      0,  &  \;\;\;\;\mbox{otherwise};
    \end{array}
   \right.  \label{eq.15}
\end{eqnarray}
\begin{equation}\label{eq.16}
\alpha_{i}^{j+1} = \frac{z^{j+1}- z^{ j }}{kz^{j+1}} + r - q-\frac{
\sigma^{2} }{2} -\frac{z^{j+1}\exp(-\xi_i)-1}{T-\tau_{j+1}}
,\;\;\beta^{j+1}=r+\frac{1}{T-\tau_{j+1}},
\end{equation}
\begin{equation}\label{eq.17}
z^{j+1}-\frac{1+r(T-\tau_{j+1})}{1+q(T-\tau_{j+1})} - \frac{\sigma^{2}
}{2}\frac{T-\tau_{j+1}}{1+q(T-\tau_{j+1})}\frac{-3y_{0}^{j+1}+4y_{1}^{j+1}-y_{2}^{j+1}}{2h} = 0\;.
\end{equation}
For the initial condition for the free boundary we have $z^{0}=\rho(0)$. An algebraic nonlinear system of equations can be derived from
(\ref{eq.14}) for $i=1,\dots,N-1$, (\ref{eq.15}) and (\ref{eq.17}). In
\cite{NielSkav} the authors apply implicit finite difference scheme,
semi-implicit scheme and upwind explicit scheme for the American put
option, combining with the penalty method. The time step parameter
for the explicit case is very small, $k=5.0\cdot 10^{-6}$.
Therefore in this work we consider the case of a fully implicit
scheme. One can also apply a scheme of the Crank-Nicolson type.

\section{Numerical Algorithms}
In order to solve the nonlinear system of algebraic equations we
developed the following two algorithms.

\smallskip
\noindent{\bf {Algorithm 1.}} This algorithm is based on the
\emph{predictor-corrector} scheme and consists in the following
steps, (see also \cite{ZhuZha,ZhuChen} for the case of pricing American put options).

{\it{Step 1.}} \emph{Predictor.} Let the solution and the free
boundary position on the time level $\tau_{j}$ be known. Instead of
(\ref{eq.17}) we use another approximation of (\ref{eq.13}) by
introducing an artificial spatial node $\xi_{-1}$:
\begin{eqnarray}\label{eq.19}
\left(1+q(T-\tau_{j+1})\right)z^{
j+1}&=&1+r(T-\tau_{j+1})+\frac{\sigma^{2} }{2} (T-\tau_{j+1})\frac{
y_{ 1 }^{ j+1 } - y_{ -1 }^{ j+1 }  } { 2h}.
\end{eqnarray}
An additional equation can be obtained from (\ref{eq.7}) by taking the
limit $\xi\rightarrow 0$ and using the fact that $\partial_\tau \Pi(0, \tau)=0$:
\begin{equation}\label{eq.20}
\alpha_{0}^{j+1} \frac{ y_{ 1 }^{ j+1 } - y_{ -1 }^{ j+1 } } { 2h} -
\frac{ \sigma^{2} }{2} \frac{ y_{ 1 }^{ j+1 } - 2 y_{ 0 }^{j+1 } +
y_{ -1 }^{ j+1 }   } { h^{2} }+\beta^{j+1} y_{ 0}^{ j+1 }= 0.
\end{equation}
Using (\ref{eq.19}) we can express $y_{-1}^{j+1}$ as:
\begin{equation}\label{eq.21}
y_{-1}^{ j+1 }=y_{1}^{j+1}- \left( qz^{j+1}- r + \frac{z^{j+1}-1}{T-\tau_{j+1}}\right) \frac{4h}{\sigma^2}.
\end{equation}
Inserting it into (\ref{eq.20}) we conclude the
following equation for the value $y_{1}^{j+1}$:
\begin{equation}\label{eq.22}
y_{1}^{j+1}=\left(\frac{2\alpha_{0}^{j+1}
h^2}{\sigma^4}+\frac{2h}{\sigma^2}
\right)\left(qz^{j+1}-r+\frac{z^{j+1}-1}{T-\tau_{j+1}}
\right)-\frac{\beta^{j+1} h^2}{\sigma^2}-1.
\end{equation}
Instead of the implicit scheme (\ref{eq.14}) we make use of its explicit variant for $i=1$ in order to derive
\begin{equation}\label{eq.23}
  \frac{ y_{ 1 }^{ j+1 } - y_{ 1 }^{ j }  }
{ k } + \alpha_{1}^{j+1} \frac{ y_{ 2 }^{ j } - y_{ 0 }^{ j } } {
2h} - \frac{ \sigma^{2} }{2} \frac{ y_{ 2 }^{ j } - 2 y_{ 1 }^{j} +
y_{ 0 }^{ j }   } { h^{2} }+\beta^{j+1} y_{ 1}^{ j}= 0.
\end{equation}
This way we obtain a nonlinear system (\ref{eq.22}), (\ref{eq.23}) for
unknowns $y_{1}^{j+1}$ and $z^{j+1}$. The system is indeed nonlinear as $\alpha^{j+1}_i$ depend on $z^{j+1}$. Now, by replacing
$y_{1}^{j+1}\leftrightarrow \widetilde{y}_{1}^{j+1}$ and
$z^{j+1}\leftrightarrow\widetilde{z}^{j+1}$ we construct the predictor value of
$\widetilde{z}^{j+1}$.

{\it{Step 2.}} \emph{Corrector}. We again use Equation (\ref{eq.14}) in a slightly
different form:
\begin{equation}\label{eq.24}
\frac{ y_{ i }^{ j+1 } - y_{ i }^{ j }  } { k } +
\widehat{\alpha}_{i}^{j+1} \frac{ y_{ i+1 }^{ j+1 } - y_{ i-1 }^{
j+1 } } { 2h} - \frac{ \sigma^{2} }{2} \frac{ y_{ i+1 }^{ j+1 } - 2
y_{ i }^{j+1 } + y_{ i-1 }^{ j+1 }   } { h^{2} }+\beta^{j+1} y_{ i
}^{ j+1 }= 0,
\end{equation}
where approximation $\widehat{\alpha}_{i}^{j+1}$ takes into account the already constructed predictor value $\widetilde{z}^{j+1}$, i.e.
\begin{equation}\label{eq.25}
\widehat{\alpha}_{i}^{j+1} = \frac{\widetilde{z}^{j+1}- z^{ j
}}{k\widetilde{z}^{j+1}} + r - q - \frac{ \sigma^{2} }{2}
-\frac{\widetilde{z}^{j+1}\exp(-\xi_i)-1}{T-\tau_{j+1}} \;.
\end{equation}
Next we use the corrected solution $y_{ i }^{ j+1 }$ and Equation
(\ref{eq.17}) in order to obtain the corrected value for the free boundary position ${z}^{j+1}$ on the next time layer.

\smallskip
\noindent{ \bf {Algorithm 2. }} We now describe an algorithm based on the
\emph{Newton method}. A variant of this method was applied for an American Call
option problem in \cite{KanVal}.

{\it{Step 1.}} We eliminate the known boundary values $y_0^{j+1}=-1$
and $y_N^{j+1}=0$ from (\ref{eq.14}). Taking into account (\ref{eq.17}) we
obtain a nonlinear system for $N$ unknowns: $y_i^{j+1}$,
$i=1,2,...,N-1$ and $z^{j+1}$. We denote by
$\stackrel{l}{\mathbf{Y}}$ the vector of these $N$ unknowns at the $l$-th
iteration.

{\it{Step 2.}} We have to solve the equation $\stackrel{l}{\mathbf{F}}=0$ with
$\stackrel{l}{\mathbf{F}}=\left(\stackrel{l}{\mathbf{F}}_{1}\;
\stackrel{l}{\mathbf{F}}_{2}\right)^T$ where $\stackrel{l}{\mathbf{F}}_{i}, i=1,2,$
correspond to Equations (\ref{eq.14}) and (\ref{eq.17}), respectively.
To this end, we apply the Newton method in the following form:
$\stackrel{l}{\mathbf{J}}(\stackrel{l+1}{\mathbf{Y}}-\stackrel{l}{\mathbf{Y})} =-\stackrel{l}{\mathbf{F}}$,
with the Jacobi matrix defined by: $\stackrel{l}{\mathbf{J}}=( \stackrel{l}{\mathbf{J}}_{ij})_{i,j=1,2}$ where
\[ \stackrel{l}{\mathbf{J}}_{11}= \left(
\begin{array}{cccccc}
     c_{1}^{j+1} & b_{1}^{j+1} &        &    &    \\
  a_{2}^{j+1} &  c_{2}^{j+1} & b_{2}^{j+1} &    &    \\
         &           \ddots  & \ddots  &  \ddots &    \\
                            &  &  a_{N-2}^{j+1} & c_{N-2}^{j+1} & b_{N-2}^{j+1} \\
                            &  &  & a_{N-1}^{j+1}  &c_{N-1}^{j+1}
\end{array}
\right), \;\; \stackrel{l}{\mathbf{J}}_{12}= \left(
\begin{array}{c}
     \frac{\partial a_{1}^{j+1}}{\partial z^{j+1}}(-1)+ \frac{\partial b_{1}^{j+1}}{\partial z^{j+1}}y_2^{j+1} \\
     \frac{\partial a_{2}^{j+1}}{\partial z^{j+1}}y_1^{j+1}+ \frac{\partial b_{2}^{j+1}}{\partial z^{j+1}}y_3^{j+1}  \\
          \vdots \\
\frac{\partial a_{N-2}^{j+1}}{\partial z^{j+1}}y_{N-3}^{j+1}+ \frac{\partial b_{N-2}^{j+1}}{\partial z^{j+1}}y_{N-1}^{j+1}      \\
\frac{\partial a_{N-1}^{j+1}}{\partial z^{j+1}}y_{N-2}^{j+1}
\end{array}
\right)
\]
$\stackrel{l}{\mathbf{J}}_{21}=
\left(\frac{-\sigma^2}{Dh},\;\;\frac{\sigma^2}{4Dh},\;\;
0,...,0\right)$ where $D=q+1/(T-\tau^{j+1})$ and
$\stackrel{l}{\mathbf{J}}_{22}=1$. Similarly
$\stackrel{l}{\mathbf{Y}}=\left(\stackrel{l}{\mathbf{Y}}_{1}
\stackrel{l}{\mathbf{Y}}_{2}\right)^T$,
$\stackrel{l}{\mathbf{Y}}_{1}=\left(y_1^{j+1},...,y_{N-1}^{j+1}\right)$,
$\stackrel{l}{\mathbf{Y}}_{2}=z^{j+1}$. As for the elements of the
matrix $\stackrel{l}{\mathbf{J}}_{11}$ we have:
\begin{eqnarray*}
  a_{i}^{j+1} &=& -\frac{1}{2h}\left(
\frac{z^{j+1}-z^{j}}{kz^{j+1}}+r-q-\frac{\sigma^2}{2}\right)-\frac{\sigma^2}{2h^2}+d_{i}^{j+1} ,\\
  c_{i}^{j+1} &=& \frac{1}{k}+\frac{\sigma^2}{h^2}+r+\frac{1}{T-\tau_{j+1}}, \\
  b_{i}^{j+1} &=& \frac{1}{2h}\left(
\frac{z^{j+1}-z^{j}}{kz^{j+1}}+r-q-\frac{\sigma^2}{2}\right)-\frac{\sigma^2}{2h^2}-d_{i}^{j+1},
\end{eqnarray*}
and
$d_{i}^{j+1}=1/(2h)(z^{j+1}\exp({-\xi_i})-1)/(T-\tau_{j+1})$.
The iteration process is repeated until the condition
$\Vert\stackrel{l+1}{\mathbf{Y}}-\stackrel{l}{\mathbf{Y}}\Vert<tol$ is fulfilled.

{\it{Step 3.}} The solution on the $(j+1)$-th time layer is
considered as an initial iteration for the next time layer. For
solving
$\stackrel{l}{\mathbf{J}}(\stackrel{l+1}{\mathbf{Y}}-\stackrel{l}{\mathbf{Y})}
=-\stackrel{l}{\mathbf{F}}$ we perform the following stages. First,
we solve the linear system of equations
$\stackrel{l}{\mathbf{J}}_{11}\stackrel{l+1}{\mathbf{Y}}_{1}=
-\stackrel{l}{\mathbf{F}}_{1}
+\stackrel{l}{\mathbf{J}}_{11}\stackrel{l}{\mathbf{Y}}_{1}
-\stackrel{l}{\mathbf{J}}_{12}\stackrel{l+1}{\mathbf{Y}}_{2}
+\stackrel{l}{\mathbf{J}}_{12}\stackrel{l}{\mathbf{Y}}_{2}$. Since
the matrix $\stackrel{l}{\mathbf{J}}_{11}$ is tridiagonal we can
apply the Thomas algorithm to find $\stackrel{l+1}{\mathbf{Y}}_{1}$.
Next, we solve
$\stackrel{l}{\mathbf{J}}_{12}\stackrel{l+1}{\mathbf{Y}}_{1}+
\stackrel{l}{\mathbf{J}}_{22}\stackrel{l+1}{\mathbf{Y}}_{2}
 =
-\stackrel{l}{\mathbf{F}}_{2}.
$
\begin{remark}
In both algorithms we choose the last time step $k-\varepsilon$ with
$\varepsilon=10^{-7}$, i.e.  $\tau_M=T-\varepsilon$. To overcome
possible numerical instabilities of these methods for
$\tau\rightarrow T$ (i.e. $t\to0$) we use the so called upwind and
downwind approximations of the term
$\frac{z^{j+1}\exp(-\xi_i)-1}{T-\tau_{j+1}}\frac{\partial\Pi}{\partial
\xi}$ depending of the sign of the term $z^{j+1}\exp(-\xi_i)-1$.
\end{remark}

\section{Numerical Experiments }
\begin{figure}[tbp]
\centering
\begin{tabular}{cc}
\makebox{\psfig{figure=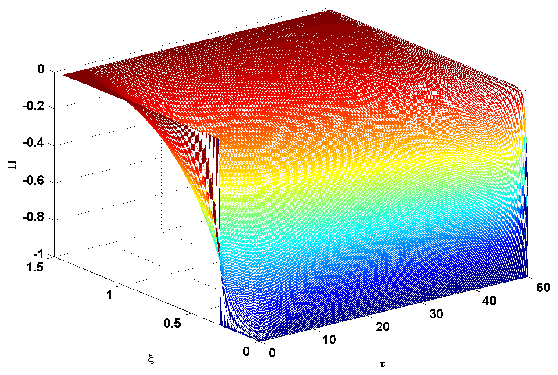,width=5cm}}
&
\makebox{\psfig{figure=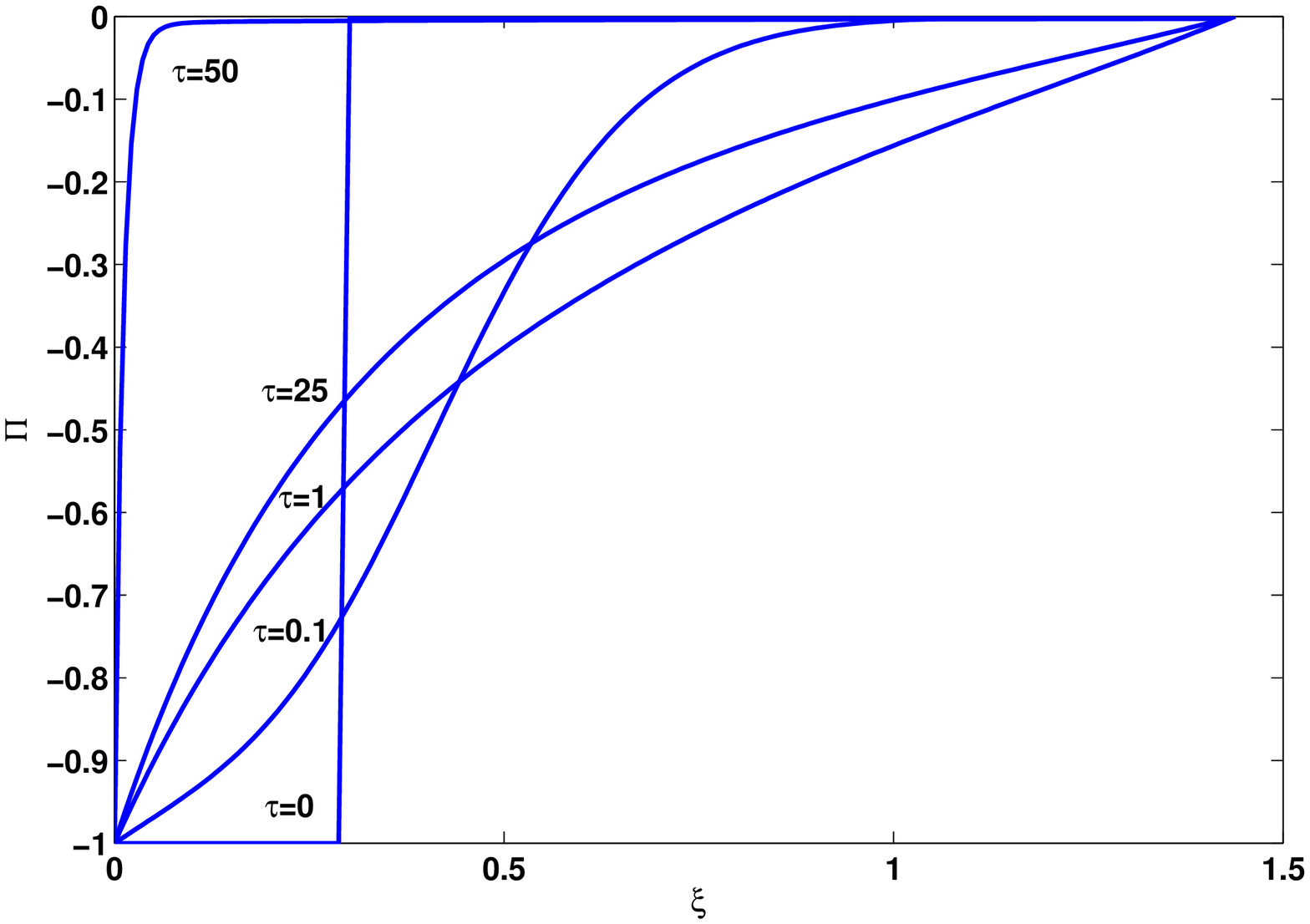,width=5cm}}
\\  a) &  b)
\end{tabular}
\caption{\label{fig.1}(a) A 3D plot of the portfolio function $\Pi$
for $T=50$, $N=200$, $M=500$; (b) Profiles of the function
$\Pi(\xi,\tau)$ for $\tau=0$, $\tau=0.1$, $\tau=10$, $\tau=25$,
$\tau=50$.}
\end{figure}

In this section we consider problem (\ref{eq.1}) with parameter
values $r=0.06$, $q=0.04$, $\sigma=0.2$ and $T=50$, taken from
examples presented in \cite{BokasSev}. Since there exists no
analytical solution to the proposed free boundary problem, we use
the mesh refinement analysis with doubling the mesh size $h$. In
Tab.~1 we present the position of the free boundary position
$\rho(\tau)$ at different times $\tau$ constructed by the Newton
method. We also present the difference between two consecutive
values and the convergence ratio are presented. The results show
nearly first order of accuracy for the free boundary and the CR
increases with increasing $\tau$ (see Tab.~1). In Fig.~\ref{fig.1}a)
a 3D plot of the portfolio function $\Pi$ for $T=50$, $N=200$,
$M=500$ is presented. In Fig.~\ref{fig.1}b) the profiles of the
function $\Pi(\xi,\tau)$ for $\tau=0,\;0.1,\;10,\;25,\;50$ obtained
by the Newton method are depicted.

In Fig. 2a) we show a comparison of the free boundary position
$\rho(\tau)$ computed by our two algorithms (Predictor-corrector and
Newton's based method) and by numerical methods from \cite{BokasSev}
(Bokes) and \cite{DaiKwok} (Kwok). It turns out that the Newton's
based method gives nearly the same results as those of
\cite{BokasSev,DaiKwok}. On the other hand, predictor-corrector
methods slightly underestimates the free boundary position
$\rho(\tau)$. In Fig. 2b) we show the free boundary position
$x_f(t)=1/\rho(T-t)$ for the original model variables $x=A/S$ and
$t$. The continuation region and exercise region are also indicated.

\begin{table}[tbp]\label{tab.1}
\caption{Mesh-refinement analysis and the convergence ratio (CR) of the Newton method.}
\scriptsize
\begin{center}
\begin{tabular}{c|c|c|c|c|c|c|c|c|c}
$N$ & $\rho(\tau=10)$& difference & CR &
$\rho(\tau=20)$ &difference& CR &$\rho(\tau=40)$ &difference& CR\\
\hline\hline 50 & 1.949988 & -          &   -    & 1.991675 & -         &  -     & 1.796663 &-&-\\
\hline 100& 1.955552 & 5.5640e-3  &   -    & 1.995525 & 3.8502e-3 &  -     & 1.803276 &6.6133e-3&-\\
\hline 200& 1.958037 & 2.4850e-3  & 1.16 & 1.996945 & 1.4194e-3 & 1.44 & 1.805149 &1.8729e-3&1.82\\
\hline 400& 1,959199 & 1.1617e-3  & 1.10 & 1.997515 & 5.7099e-4 & 1.31 & 1.805667 &5.1799e-4&1.85\\
\hline 800& 1.959758 & 5.5965e-4  & 1.05 & 1.997765 & 2.4919e-4 & 1.20 & 1.805813 &1.4621e-4&1.82\\
\end{tabular}
\end{center}
\end{table}

\section{Conclusions} In this paper we have analyzed numerical algorithms for solving
the free boundary value problem for American style of floating
strike Asian options. To solve corresponding degenerate parabolic
problem we have applied Landau's front fixing  transformation
method. We proposed two numerical algorithms based on the
predictor-corrector scheme and the Newton's method. The
predictor-corrector scheme is computationally faster when compared
to the Newton method. It yields a good approximation close to
expiry. However, its accuracy is decreased for times close to the
initial time. The second algorithm based on Newton's method yields
better approximation results over the whole time interval. Although
all finite difference approximations are of second order, due to
discontinuity of the initial datum and nonlinear behavior of the
coefficients in all discrete equations, the results show nearly the
first order rate of convergence.

\section*{Acknowledgments}
The first author was supported by projects  Bg-Sk-203/2008 and DID 02/37-2009. The second author was supported by the project APVV SK-BG-0034-08.
\begin{figure}[tbp]
\centering
\begin{tabular}{cc}
\makebox{\psfig{figure=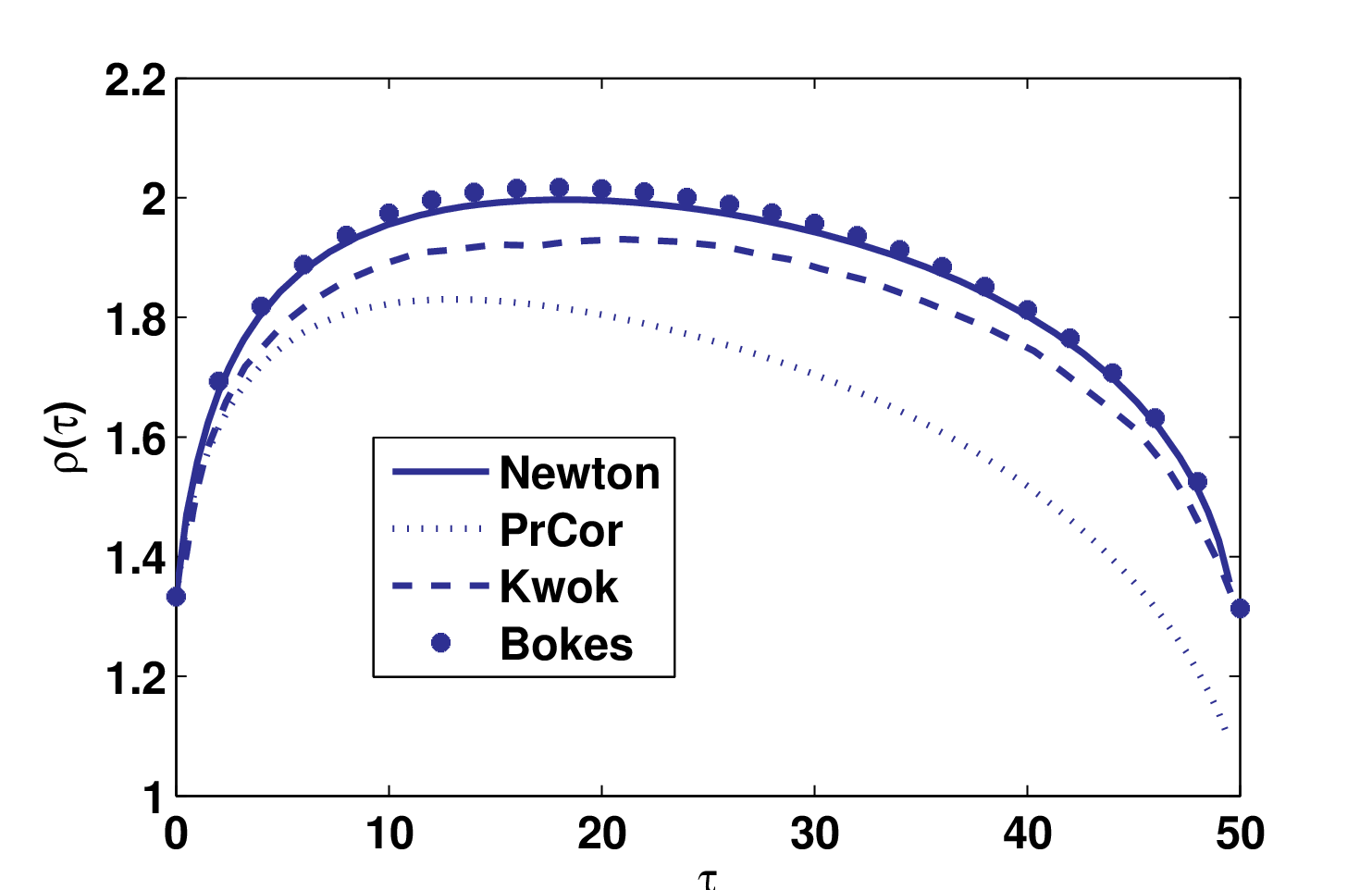,width=6.2cm, height=3.5cm}}&
\makebox{\psfig{figure=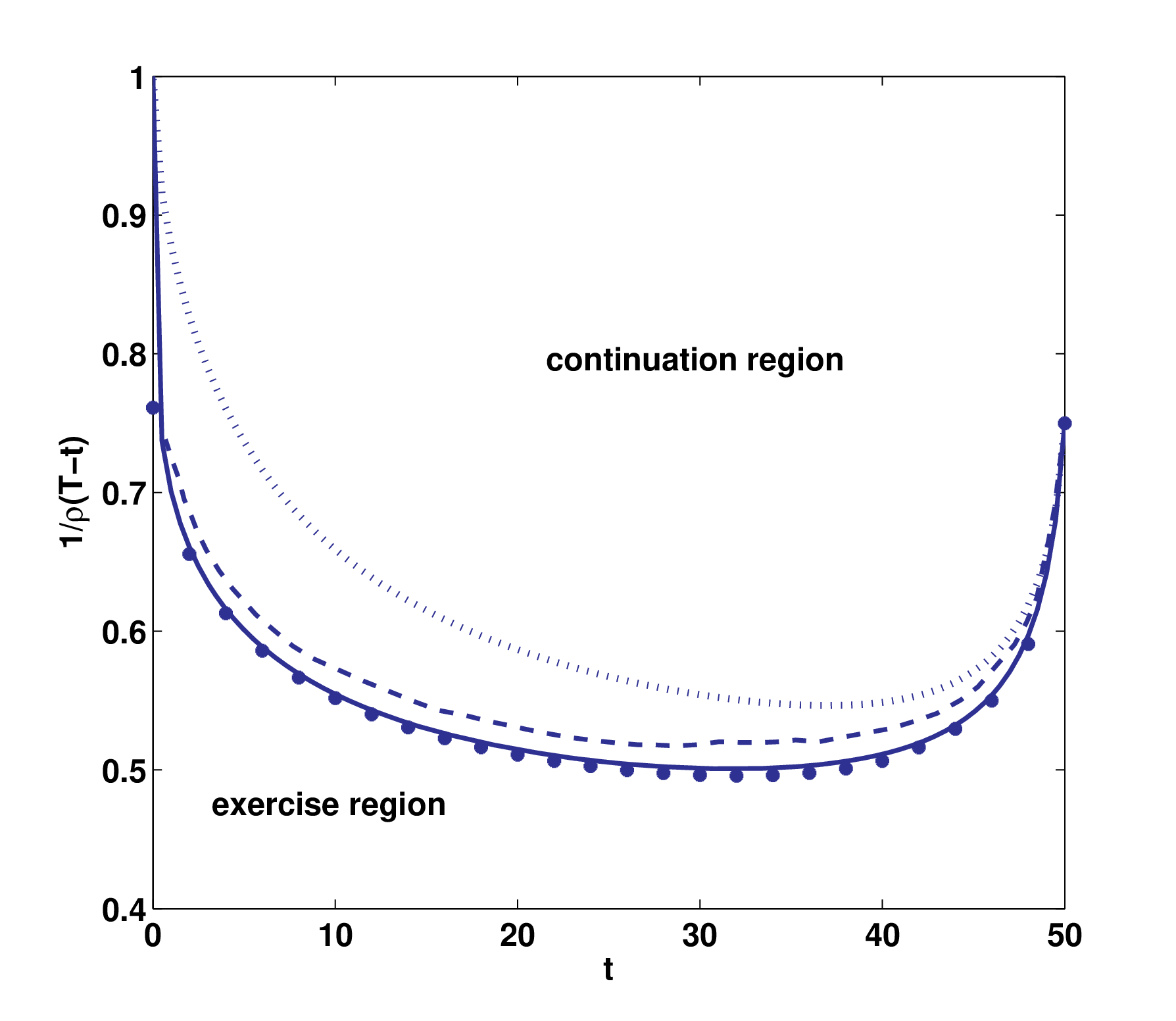,width=6.2cm, height=3.5cm}}
\\  a) &  b)
\end{tabular}
\caption{a) Comparison of the free boundary $\rho(\tau)$ for various methods; b) the
free boundary position $x_f(t)=1/\rho(T-t)$ splitting the continuation and
exercise region of American style of Asian call option.}
\end{figure}


\begin{thebibliography}{99}
\bibitem{BokasSev} Bokes, T., \v{S}ev\v{c}ovi\v{c}, D.: Early exercise boundary for American type of floating strike Asian option and its numerical
approximation, to appear in: Applied Mathematical Finance, 2011.


\bibitem{DaiKwok} Dai, M., Kwok, Y.K.: Characterization of optimal
stopping regions  of American Asian and lookback  options. Math.
Finance 16(1) (2006) 63--82.



\bibitem{Gupta}
Gupta, S. C.: The Classical Stefan Problem: Basic Concepts,
Modelling and Analysis. North-Holland Series in Applied Mathematics
and Mechanics, Elsevier, Amsterdam (2003).

\bibitem{HanWu} Han, H., Wu X.: A fast numerical method for the
Black-Scholes equation of American options. SIAM J. Numer. Anal.
41(6) (2003) 2081--2095.

\bibitem{KanVal}  Kandilarov, J., Valkov, R.:  A Numerical Approach
for the American Call Option Pricing Model, Lecture Notes in
Computer Science 6046 (2011) 453--460.

\bibitem{Kwok}  Kwok., J. K.: Mathematical Models of Financial Derivatives.
Springer-Verlag (1998).

\bibitem{LauSev} Lauko, M.,  \v{S}ev\v{c}ovi\v{c}, D.: Comparison of numerical and analytical approximations of the early exercise boundary of the American put option. ANZIAM journal 51 (2010) 430--448.


\bibitem{MoyScar} Moyano, E., Scarpenttini, A.: Numerical stability
study and error estimation for two implicit schemes in a moving
boundary problem. Num. Meth. Partial Differential Equations, 16(1)
(2000) 42--61.

\bibitem{NielSkav} Nielsen, B., Skavhaug, O., Tveito, A.: Penalty
and front-fixing methods for the numerical solution of American
option problems, Journal of Comput. Finance, 5(4) (2002) 69--97.



\bibitem{Sev} \v{S}ev\v{c}ovi\v{c}, D.: Analysis of the free boundary for the
pricing of an  American call option. Eur. J. Appl. Math. 12 (2001) 25--37.

\bibitem{SevcovEhrh}  \v{S}ev\v{c}ovi\v{c}, D.: Transformation methods for evaluating approximations to the optimal exercise boundary for linear and
nonlinear Black-Sholes equations. In: M. Ehrhard (ed.), Nonlinear
Models in Mathematical Finance: New Research Trends in Optimal
pricing, Nova Sci. Publ., New York (2008) 153--198.

\bibitem{SevTak} \v{S}ev\v{c}ovi\v{c}, D., Tak\'a\v{c}, M.: Sensitivity analysis of the early exercise boundary for American style of Asian options, to appear in: International Journal of Numerical Analysis and Modeling, Ser. B, 2011.


\bibitem{StaSev}
Stamicar, R., \v{S}ev\v{c}ovi\v{c}, D., Chadam, J.: The early exercise boundary
for the American put near expiry: Numerical approximation. Canadian
Applied Mathematics Quarterly, 7(4) (1999) 427--444.

\bibitem{Tangman}  Tangman, D. Y.,  Gopaul, A., Bhuruth, M.:
A fast high-order finite difference algorithms for pricing American
options. J. Comp. Appl. Math. 222 (2008) 17-29.



\bibitem{ZhuZha}  Zhu, S.-P.,  Zang, J.: A new predictor-corrector scheme for valuing American puts. Applied Mathematics and Computation 217 (2011)  4439--4452.

\bibitem{ZhuChen}  Zhu, S.-P.,  Chen, Wen-Ting:
A predictor�corrector scheme based on the ADI method for pricing American puts with stochastic volatility. to appear in: Computers \& Mathematics (2011)


\end{thebibliography}
\end{document}